\documentclass[aps,preprintnumbers,superscriptaddress,showpacs,twocolumn,longbibliography]{revtex4}

\usepackage{epsfig}
\usepackage{psfrag}
\usepackage{amsfonts}
\usepackage{amsmath}
\usepackage{graphicx}
\usepackage{float}
\usepackage{subfig}
\usepackage{dcolumn}
\usepackage{bm}

\begin{document}
\title{ The evolution of Information entropy components in Relativistic Heavy-Ion Collisions  }

\author{Fei Li}
\affiliation{School of mathematics and
physics, China University of Geosciences(Wuhan), Wuhan 430074,
China}

\author{Gang Chen}
\email{chengang1@cug.edu.cn} \affiliation{School of mathematics and
physics, China University of Geosciences(Wuhan), Wuhan 430074,
China}

\begin{abstract}
Shannon information entropy provides an effective tool to study the evolution process in relativistic
heavy-ion collisions. The time evolution process of thermodynamic entropy $S_{\rm thermal}$, multiple
entropy $S_{\rm mul}$, and configuration entropy $S_{\rm conf}$ at RHIC is studied using the AMPT model
to generate central Au-Au collisions. By superimposing the three kinds of information entropy, we can
get a more complete information entropy of the system to describe the physical information of the relativistic
heavy ion collision. The results show that the four stages of the time evolution process of the system
entropy $S$ seem to correspond to the four physical processes in the relativistic heavy ion collision,
indicating that the total entropy of the system can reflect the physical information in the relativistic
heavy ion collision more accurately.

\end{abstract}


\maketitle
\section{Introduction}
Search for clear signatures of the phase transition in Relativistic heavy-ion
collisions (RHIC) is an important subject in high-energy physics.
The process of phase transition in collisions at RHIC includes
the transformation of quark-gluon plasma (QGP) into hadron gas.
Physicists believe that this process reproduces the situation
of the first 10¦Ìs after the Big Bang~\cite{Turko2018}. Further study of phase
transition can bring a deeper understanding of the properties of
the nuclear matter produced in the interaction of heavy nuclei
and the mechanism of the origin of the universe.
However, due to the complexity of the dynamic properties of the phase
transition process in heavy-ion collisions, there is no unified theory
to describe the whole reaction process at present. In this dilemma, researchers
use different theoretical approaches to describe different stages in nuclear collisions~\cite{Werner2001Tools},
and some researchers simultaneously devoted themselves to the study of entropy generated in collision process
by using different dynamic models in the 1980s~\cite{Csernai1986Entropy}. In recent studies,
Shannon information entropy has been adopted and developed by some
researchers because it provides a new method and observation~\cite{Ma2018}.

Based on the works~\cite{boltzmann2012lectures,Boltzmann2003,jaynes1965gibbs,cropper2001great}
of Boltzmann and Gibbs in the last 20 years of the 19th century,
today we understand that the entropy of the system can be determined by its specific
probability distribution $p_{i}$, for a system in state \textit{i}. In 1948, C.E. Shannon discovered a
theorem similar to the definition of ``entropy" in physics, which can be expressed as
follows~\cite{Shannon1948,longair1984theoretical}:
\begin{equation}
S=-k_{B}\sum_{i} p_{i} ln p_{i}
\end{equation}
where $p_{i}$ (i=1, 2, \dots, n) is the independent probabilities of events in a
system and $k_{B}$ is Boltzmann¡¯s constant. For a given set of constraints, when $p_{i}$ is the
most probable state of the system, the information entropy has a maximum value.
In terminology of physics, when the system is in the most probable distribution, its
entropy is the maximum, which corresponds to its equilibrium state. It should be noted
that the above definition of entropy is very broad, which can be used not only for the
equilibrium state of the system, but also for the non-equilibrium state of the system.

The Shannon information entropy has been applied and developed in many scientific areas~\cite{Ma2018}.
In the area of heavy-ion collisions, Cao and Hwa first applied Shannon information entropy
to the study of chaotic behavior caused by particle production in branching process~\cite{cao1996chaotic}.
In 1996. Y.G. Ma adopted the idea of ``event entropy" to obtain a novel signature of liquid
gas phase transition under the references of transition temperature in 1999~\cite{ma1999application}. In the
study of intermediate mass fragments, C.W. Ma et al. adopted the event information entropy
found the isobaric scaling phenomenon in neutron-rich projectile fragmentation reactions~\cite{Ma2015},
as well as, in the fragments differing of different neutron-excesses~\cite{ma2016scaling}. Recently, J. Xu and
C.M. Ko investigated the chemical freeze-out conditions in RHIC by specific entropy of hadrons~\cite{Xu2017}.

In the process at RHIC, the scattering between particles, the production and annihilation
of particles, and the variation of the internal structure of particles are always accompanied.
Because information entropy can be expressed by different stochastic variables in different physical conditions,
it is suitable to adopt the thermodynamic entropy, multiplicity entropy, and configuration
entropy to describe these different types of information changes. Here, the distribution $\{p_i\}$ of
thermodynamic entropy is defined by the distribution of 6-dimensional phase space of particles,
which measures the disorder degree of particles in phase space, and the distribution $\{p_i\}$ of
multiplicity entropy is defined by the event probability of having ¡°\textit{i}¡± particles produced, which
measures the disorder of particles in event phase. The adoption of configuration entropy in this
work is inspired by Csernai et al.~\cite{Csernai2017} and Lichtenberg~\cite{Lichtenberg1982}, makes us consider the information variables generated by the different composition of quarks inside hadrons, because hadrons are not a
point particle. The introduction of these three different entropies will be detailed in the
following section.

The organization of the present letter is as follows. In section 2, The A multiphase transport model (AMPT),
which we used in this work, will be briefly introduced.
In section 3 and 4, the thermodynamic entropy and multiplicity entropy in the evolution of heavy-ion
collisions is investigated. In section 5, the configuration entropy of hadron internal structure will
be introduced, the configuration entropy and the total information entropy in the evolution of heavy-ion
collisions is investigated as well.

\section{The AMPT Model}
A multiphase transport model~\cite{Lin2005} is used to analyze
the evolution of information entropy, which is a widely used theoretical tool for relativistic heavy
ion collisions. The AMPT model is based on nonequilibrium transport dynamics,
which consists of four parts: the Heavy-Ion Jet INteraction Generator (HIJING)
model~\cite{Wang1991} for generating the initial-state information, Zhang¡¯s parton cascade
(ZPC) model~\cite{Zhang1998} for modeling partonic scatterings, the Lund string fragmentation
model or a quark coalescence model for hadrons formation, and a relativistic
transport (ART) model~\cite{Li1995} for treating the resulting hadron scatterings. These are
combined to give a coherent description of the dynamics of relativistic heavy
ion collisions.
\begin{figure}[htp]
\centering
\includegraphics[width=7.0cm]{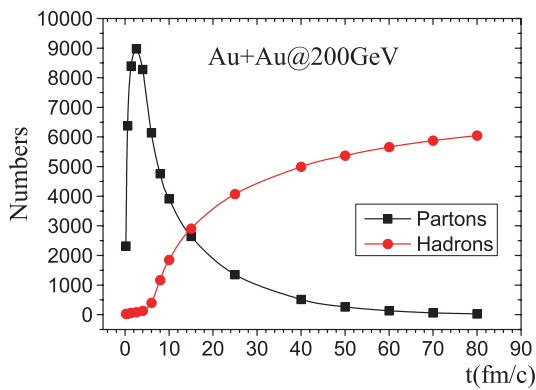}
\caption{ Time evolution of the parton and hadron production in the central
 Au+Au collisions at $\sqrt{s_{NN}}$ = 200~GeV calculated by AMPT.}
\label{label}
\end{figure}

Here we choose the string melting version of the AMPT model (v2.26t9b;isoft=5),
in which partons freeze-out according to local energy density. The hadronization
process is realized by a quark cascade model, which combines two nearest
partons into a meson and three nearest quarks (anti-quarks) into a baryon
(anti-baryon). The method of determining hadron species is done by the flavor
and invariant mass of coalescing partons. The impact parameter is in
the range b $\leq$ 3~fm and the parton cross section is taken to be 10~mb. To
give a general picture of the evolution of particle production after collision,
we obtain the time evolution of the number of partons and hadrons in the
central Au+Au collision at the $\sqrt{s_{NN}} = 200$~GeV by
using AMPT. As shown in Fig.~1, a large number of partons are produced in
the early stage of collision $(t< 5$~fm/c), while later only a few hadrons are produced.
During these times, partons are dominant, and the whole system is in a deconfined
phase, which is in the perturbative QCD vacuum, with only a few of hadrons~\cite{Meiling2006}.
After that, the number of hadrons increases with the decrease
of the number of partons, and the system goes through a phase transition into hadron gas phase.

\section{The Thermodynamic entropy}
The thermodynamical entropy can be calculated by the 6-dimenssional
phase space distribution $p_{i}(p_{x}, p_{y}, p_{z}, x, y, z)$.
Under the definition of shannon entropy, the $p_{i}$ is the ratio of the particle
number in the local 6-dimensional phase-space i.e. the $i$-th bin in the global 6-dimensional
phase space. Here, the number and size of 6-dimensional
phase space units are set to sufficiently contain all the mechanical information after the
average of all events. It should be noted that the distribution of the system is based on the
number of certain particles versus the total number of particles. The thermodynamic entropies
of the partons, hadrons, and system are calculated by Eq.~(1) in Au+Au collisions at
$\sqrt {s_{NN}}= 64.2$~GeV, 200~GeV, and 800~GeV.


\begin{figure}[htp]
\centering
\includegraphics[width=6.0cm]{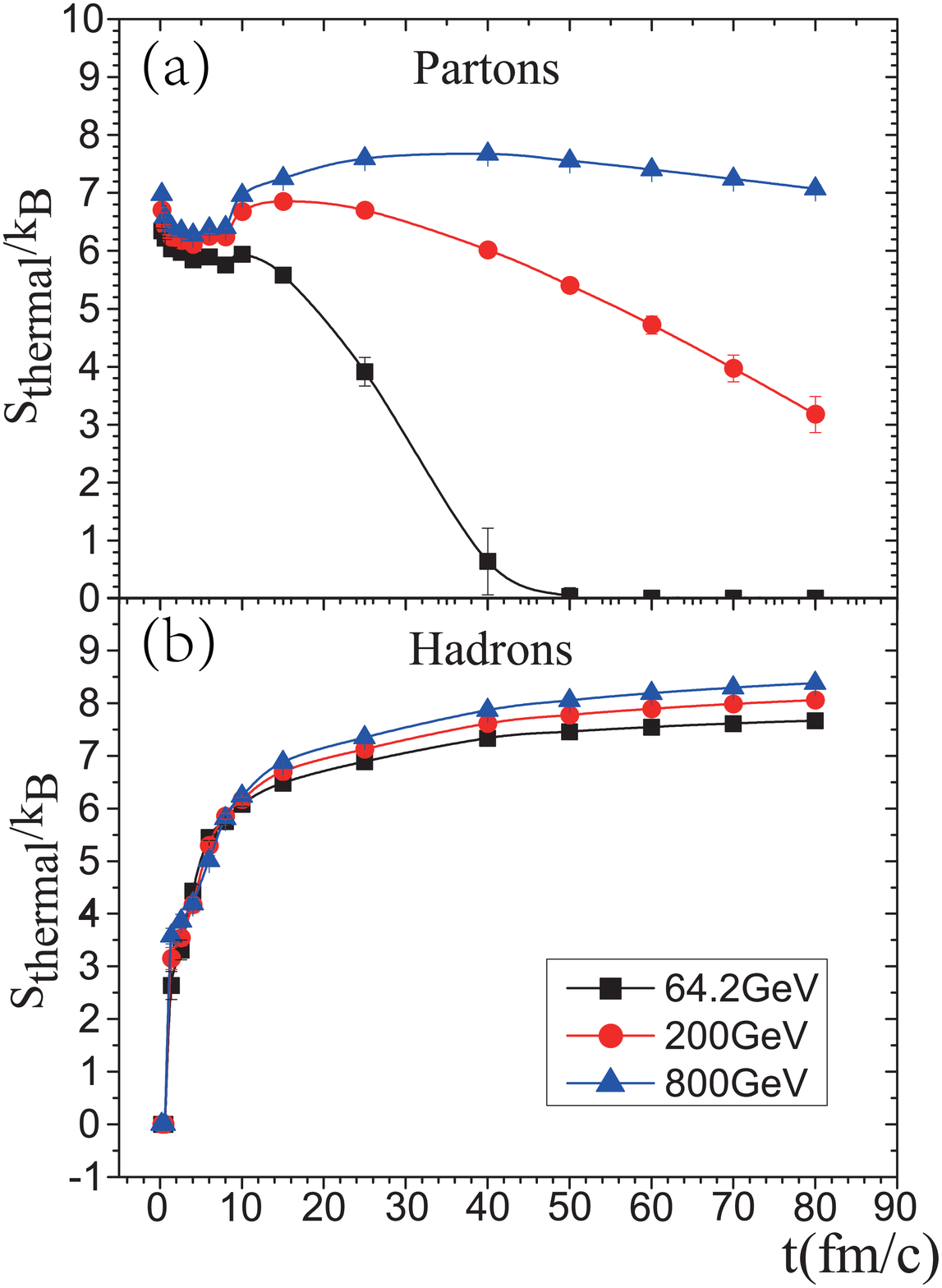}  
\caption{ Time evolution of thermodynamic entropy in Au+Au collisions at different c.m. energies for (a) partons and (b)  hadrons.}
\label{label}
\end{figure}

We can be seen from Fig.~2(a), that the thermodynamic entropy of partons at
different center of mass of energies (c.m. energy) have similar evolution curves. The trend of the
thermodynamic entropy decreases first, then goes up and reaches a vertex,
and then decreases. The reason for the decrease of parton thermodynamic
entropy in the first stage is that strings are not included in our statistics,
and then the energy released by the string melting is used as the external energy
input to the system.
The increase of the thermodynamic entropy of partons
in the second stage is due to the scattering and generation of partons. The
decrease of thermodynamic entropy of partons in the third stage is due to the
disappearance of cooling of partons. It can be observed that the thermodynamic
entropy of the final parton disappears earlier in low-energy collisions because
the central local temperature is lower in low-energy collisions, so the partons
freeze out faster.

From Fig.~2(b), the thermodynamic entropy of all hadrons increase rapidly in the early
stage at three different c.m. energies, due to the rapid generation of hadrons. Then, the
thermodynamic entropy of hadrons gradually tends to saturation and reaches a maximum,
which means that the system tends to equilibrium.

\begin{figure}[htp]
\centering
\includegraphics[width=7.0cm]{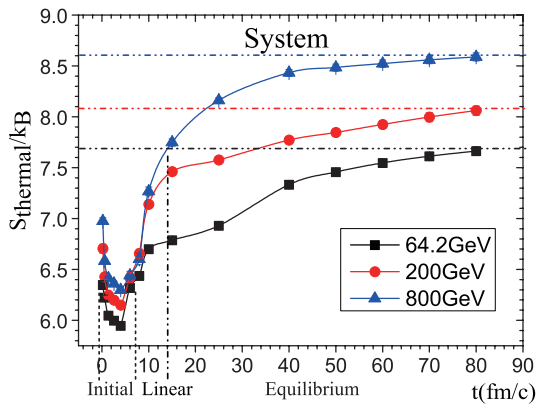}
\caption{ Time evolution of thermodynamic entropy for system in Au+Au collisions at different c.m. energies.}
\label{label}
\end{figure}

Fig.~3 shows the time evolution of thermodynamic entropy for the whole collision system,
which goes through three stages. The first stage is the initial state of parton, its total
thermodynamic entropy decreases and increases due to the energy released and gained by string formation and melting.
It also can be interpreted in physical images as entropy reduction caused by compression and expansion
after collision. At the same time, the quark gluon plasma (QGP) is formed when the entropy
decreases to the minimum.
In the second stage, the thermodynamic entropy linearly increases at
$t=5$ to 10~fm/c, which corresponds to the phase transition process of the system from
QGP to hadronic gas. The third stage is the process of thermodynamic entropy approaching saturation,
which corresponds to the final equilibrium state of hadron scattering. In addition, the higher
the c.m. energy in the high-energy collision, the faster the thermodynamic entropy of the system
reaches the equilibrium state. This is because the particles produced by collisions with higher
energies have higher momentum, which makes the system approach the maximum disorder faster.

\section{The Multiplicity entropy}

Above we studied the time evolution of thermodynamic entropy in high-energy collision systems.
In fact, there is also a multiplicity entropy in high-energy heavy ion collisions.
In 1999, Y.G. Ma introduced this method~\cite{ma1999application} to diagnose a nuclear liquid gas
phase transition by multiplicity entropy, which determines the critical point by finding the
maximum value of multiplicity entropy in a certain state of the system. Here, in the
context of multiplicity entropy, the probability distribution $p_{i}$ is related to the
ratio of the particle numbers $N_i$ produced in the $i$-th bin to the total particle number $N$,
i.e. $p_{i} = N_{i}/N$. ${p_{i}}$ is the normalized probability distribution,
where $\sum _{i}p_{i} = 1$. The time evolution of the multiplicity entropy of partons and
hadrons at different c.m. energies is calculated by Eq.~(1), as shown in Fig.~4.

\begin{figure}[htp]
\centering
\includegraphics[width=6.5cm,height=10cm]{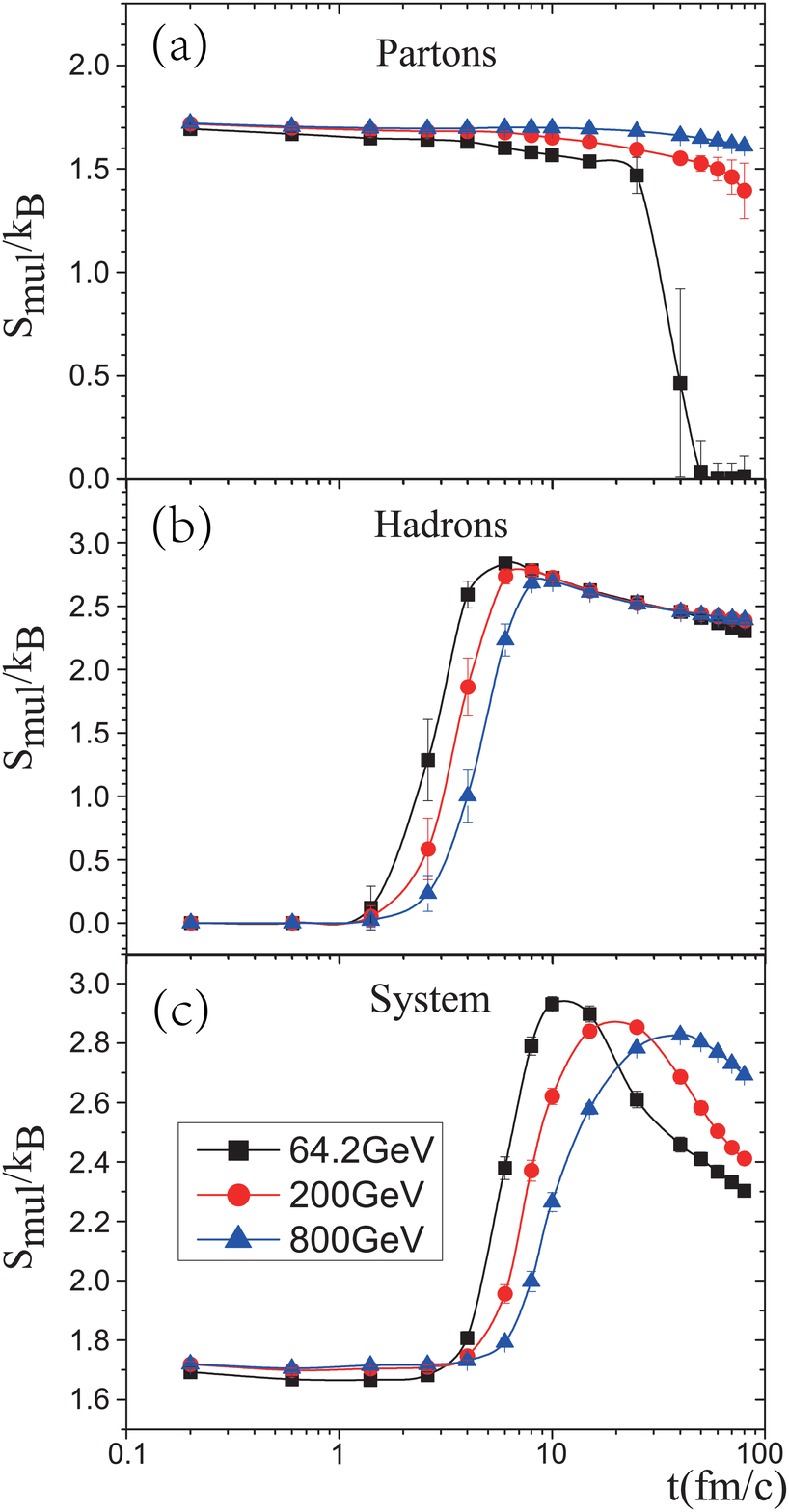}
\caption{ Time evolution of multiplicity entropy in Au+Au collisions at different c.m. energies for (a) partons, (b) hadrons, and (c) system.}
\label{label}
\end{figure}

From Fig.~4(a), we can see that the time distribution of multiplicity entropy of partons are all similar under
different c.m. energies of Au-Au collisions. The multiplicity
entropy of partons decreases slowly before 10~fm/c of the evolution process,
which is due to the fact that the distribution of various kinds of partons
remains almost unchanged at this stage. Under the conditions of $\sqrt {s_{NN}}=200$~GeV and
800~GeV, the multiplicity entropy of partons decreases slightly after 10 fm/c,
which is due to the faster cooling of energetic quarks at the later stage
of the collision system. Under the condition of $\sqrt {s_{NN}}=800$~GeV, the change of multiplicity
entropy is smaller than that of $\sqrt {s_{NN}}=200$~GeV, because the energetic quarks can exist
longer under higher energy collisions, which has little effect on the overall
distribution of partons. At $\sqrt {s_{NN}}=64.2$~GeV collisions, the multiplicity
entropy drops sharply near 30~fm/c because the partons almost completely freeze out.

In Fig.~4(b), a significant inflection point appears about 6-8~fm/c of hadronic
multiplicity entropy, which is the maximum of multiplicity entropy at different c.m. energies, This seems
to imply the critical point of the system from QGP to hadronic gas. This inflection point gives a good sign
of phase transition because the maximum of the multiplicity entropy reflects the largest fluctuation
of the multiplicity probability distribution at the critical point. From
the perspective of information theory, the prediction of which hadron will appear
in the system at this moment is the most difficult. Fig.~5(a) shows that the maximum value of multiplicity
entropy appears at about 7~fm/c, and the corresponding temperature in Fig~5(b) at the same time is 147 MeV,
which is almost the same as the chemical freeze-out temperature of 141~MeV obtained by specific entropy
in reference~\cite{Xu2017}. Both results are slightly lower than those extracted from the
experimental data based on the statical model~\cite{Adams2005,Adcox2005,Adamczyk2017,Cleymans2006,Andronic2010}.
\begin{figure}[htp]
\centering
\includegraphics[width=8.0cm]{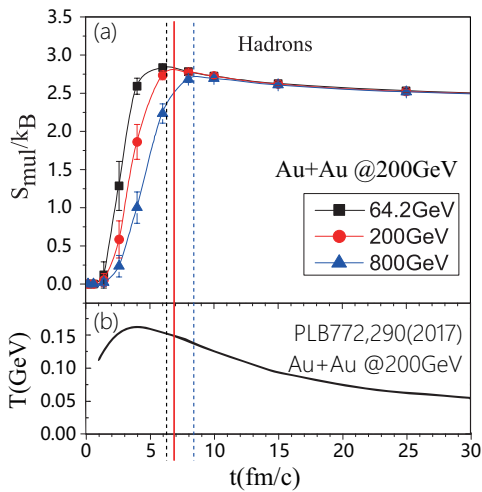}
\caption{(a) Time evolution of the Multiplicity entropy in the hadronic phase of central Au+Au
collisions at $\sqrt{S_{NN}}$ = 200 GeV, as well as, (b) the extracted temperature from the hadron
resonance gas model.}
\label{label}
\end{figure}

It should be noted that the critical point obtained by the multiplicity
entropy is very close to the starting point of the linear increase of the thermodynamic
entropy, which indicates that the chaotic degree of the system increases sharply after
the system evaporates from QGP state to hadronic gas.
The critical point will be reached later at higher c.m. energies, due to the higher temperature
in the central region at higher c.m. energies.
Thus the cooling time of QGP into hadronic gas will be longer. After the critical point,
the multiplicity entropy of hadrons decreases gradually to a stable value, which
reflects the decay of unstable excited hadrons into more stable hadrons.

The multiplicity entropy of the whole system is also plotted in Fig.~4(c), but this result
is unsatisfactory for the prediction of critical point. The reason for this defect
is that the particles in the global region can not correspond to those particles in the region
where the phase transition actually occurs, while the region where the hadron is produced
corresponds to the region where the phase transition occurs. Therefore, it is more accurate
to determine the critical point by using the hadron multiplicity entropy.

\section{The configuration entropy}
In the previous studies, we regard both partons and hadrons as point particles without
internal structure, but in fact mesons and baryons are composed of quarks, in which the
mesons are composed of two quarks and most baryons are composed of three quarks. For
mesons consisting of two quarks, there is only one space configuration, but for baryons
consisting of three quarks, the internal space configuration is not clear. It should be
noted here that after nuclear collision, the system will inevitably be accompanied by
changes in the amount of information in the process from point particles without internal
structure to nucleons with internal structure. Here we attempt to quantify the information
variable caused by the change of particle internal configuration.

Inspired by the work of Csernai et al.~\cite{Csernai2017}, we consider that particles consisting of three nucleons
have four possible configurations, each of which has a different probability of formation,
which is related to: direction dependence of the links, different constituents, different
(energetic) weights of the links, dynamical freedom of the length or angle of the link, etc.
In this way, we can get the corresponding topological Shannon entropy through the probability
distribution of different binding modes of quarks in nucleon.

\begin{figure}[htp]
\centering
\includegraphics[width=3.2cm]{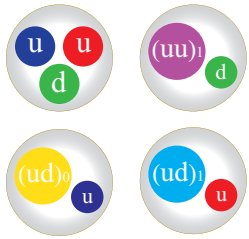}
\caption{ The 4 internal configurations of proton based on the Di-quark model.}
\label{label}
\end{figure}

It is interesting that the quark-diquark model mentioned by Lichtenberg~\cite{Lichtenberg1982, Rapp1998,Tanabashi2018} in 1982 coincides with the above viewpoint. This model described baryon as a bound state of one quark with one
diquark, in which the diquark is formed by the attraction of two quarks with color and spin
anti-symmetry, when both quarks are correlated in this way they tend to form a very low energy
configuration. In this model, there are four possible configurations inside baryons (we
only consider those particles with quark numbers less than or equal to 3).
For example, the four internal configurations of protons are shown in Fig.~6. Here we assume
equal probability for each topological configuration in baryon, so that the probability of any
configuration in baryons is 1/4. However, we also need to consider that there are other particles
with only one internal configuration in the collision process, so there are:
\begin{small}
\begin{equation}
p_{\rm baryon} = \frac{N_{\rm baryon}}{N} \quad ,\quad p_{\rm{other}} =  1 - p_{\rm baryon}.
\end{equation}
\end{small}

Then the configuration entropy of system is:
\begin{small}
\begin{equation}
\begin{split}
S_{\textrm {conf}} &= (p_{\textrm {other}})ln(p_{\textrm {other}})+
4*[\frac{p_{\textrm {baryon}}}{4}ln(\frac{p_{\textrm {baryon}}}{4})].\\
\end{split}
\end{equation}
\end{small}

\begin{figure}[htp]
\centering
\includegraphics[width=7.0cm]{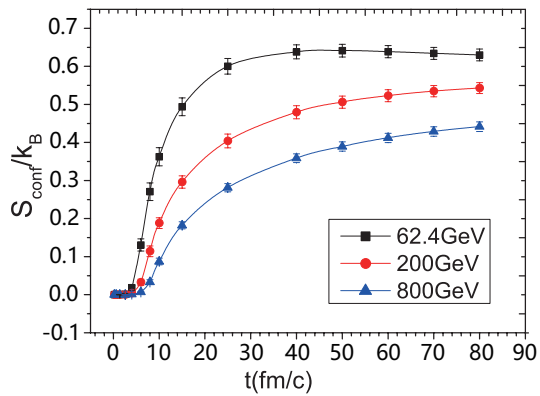}
\caption{ Time evolution of configuration entropy of system in Au+Au collisions at different energies.}
\label{label}
\end{figure}

The configuration entropy of system in Au+Au collisions at different c.m. energies is plotted in
Fig.~7. One can see that the time evolution of configuration entropy at different c.m. energies are
similar in Fig.~7. The configuration entropy of a collision system with a higher c.m. energy is
lower than that of a collision system with a lower c.m. energy,
because higher c.m. energies collide to produce a larger proportion of mesons without internal structures.

Above all, we have calculated the thermodynamic entropy, multiplicity entropy, and configuration
entropy of a Au+Au collision system at different c.m. energies of 64.2~GeV, 200~GeV, and 800~GeV,
respectively. Now we need to rethink that what characteristics of a Au+Au collisions at the RHIC
energy are showing by these different kinds of entropy.

\begin{figure*}[htp]
\centering
\includegraphics[width=16.2cm]{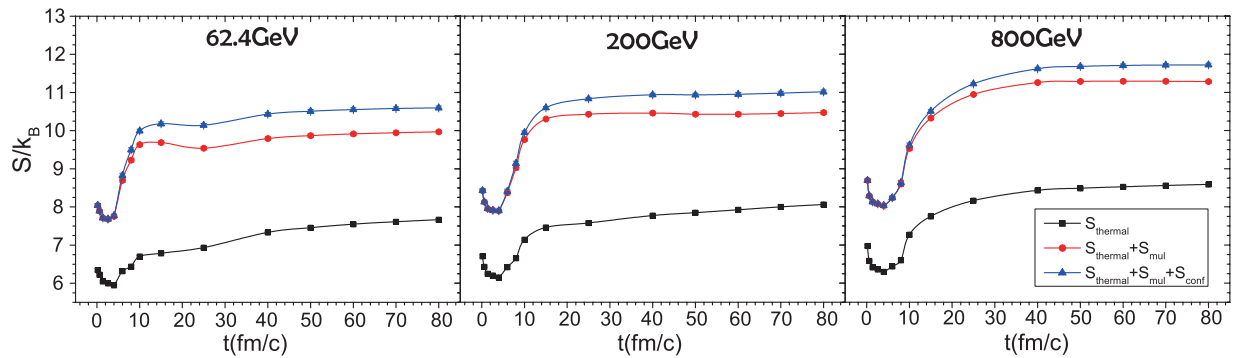}
\caption{ Time evolution of total information entropy for system in Au-Au collisions at different c.m. energies.}
\label{label}
\end{figure*}

In C.W. Ma's review~\cite{Ma2018}, ``the basic scientific meaning of Shannon information entropy
applied in heavy-ion collisions is to indicate the chaoticity of nuclear matter in the colliding
nuclear system" is mentioned. This reminds us that all three kinds of entropy, including from dynamic
phase space, particle type and internal configuration of particles, reflect the chaotic nature of the
system in the evolution of relativistic heavy ion collisions. This enlightens us that a complete
description of the degree of chaos in the evolution of relativistic heavy ion collision system should
include as much information entropy changes as possible in different aspects. The cumulative values
of the above three kinds of entropy are calculated under this idea. The results of the total entropy
$S$ ($=S_{\rm thermal}+S_{\rm mul}+S_{\rm conf}$) of the system during the evolution of Au+Au
collisions are shown in Fig.~8, where the time development appear smooth.

From Fig.~8, we can see that the time evolution of the total entropy in the Au+Au collisions system
goes through four stages: The first stage, about $0<t<3$~fm, entropy decreases due to the energy
released by string formation and melting and compression after collision; Second, entropy is at the
minimum of the system about $3<t<5$~fm, corresponding to the parton rescattering phase, which the
system is in the quark gluon plasma (QGP) state; Then the entropy of the system increases rapidly
from $t=5$~fm to 9~fm, which may correspond to the transformation process from QGP to hadron phase;
After that, the entropy distribution tends to be saturated, which would correspond to state of hadron
scattering. In addition, the higher the c.m. energy in the Au+Au collision, the faster the entropy of
the system reaches the equilibrium state. This is because the particles produced by collisions with
higher energies have higher momentum, which makes the system approach the maximum disorder faster.

Following the reference~\cite{Csernai2017}, it is important to mention that, (1) all degrees of freedom should
be taken into account, (2) these should be independent degrees of freedom (orthogonal),
(3) the degrees of freedom must be quantized, also for continuous degrees of freedom. This last point
was recognized after the development of quantum mechanics, which led to the quantization of phase-space
volume cells. Ponit~(1) is violated in this AMPT evaluation on the string degrees of freedom are not
taken into account and this led to a temporary decrease of the total entropy. Such decrease should not happen
in a closed system. This missing entropy of string is shown clearly in Fig.~8
~\\

In this letter, the time evolution process of thermodynamic entropy $S_{\rm thermal}$, multiple entropy
$S_{\rm mul}$, and configuration entropy $S_{\rm conf}$ in relativistic heavy ion collision is studied
carefully using AMPT model to generate central Au+Au collisions at $\sqrt {s_{NN}} = 200$~GeV with $ |y| < 1$
and $p_T < 5$ acceptances. In this way, we obtain the time evolution distribution image of the total entropy
$S$ in the relativistic heavy ion collision. The results show that the four stages of the time evolution
process of the system entropy $S$ have obvious correspondence with the four physical processes experienced
in the relativistic heavy ion collision, indicating that the total entropy of the system can more accurately
reflect the physical information in the relativistic heavy ion collision.

\acknowledgments{The authors thank Prof. Laszlo P. Csernai and Prof. Che Ming Ko for valuable comments, thank
Ph.D. L. Zheng, Z.L. She, and Z.J. Dong for useful discussion about AMPT. This work is supported by the
National Natural Science Foundation of China(11475149).
\section{References}
\bibliography{entropy}

\end{document}